\documentclass[11pt]{article}
\usepackage{moriond,epsfig}

\def\be{\begin{equation}}
\def\ee{\end{equation}}
\def\bea{\begin{eqnarray}}
\def\eea{\end{eqnarray}}

\begin{document}
\title{COULOMB EFFECTS AND ELECTRON TRANSPORT THROUGH A COHERENT CONDUCTOR}

\author{\underline{A.D. ZAIKIN}$^{1,2}$,
D.S. GOLUBEV$^{2,3}$}

\address{$^1$Forschungszentrum
Karlsruhe, Institut f{\"u}r Nanotechnologie, D-76021 Karlsruhe,
Germany}
\address{$^2$I.E.Tamm Department of Theoretical Physics, 
P.N.Lebedev Physics
Institute, Leninskii pr. 53, 117924 Moscow, Russia}
\address{$^3$Institut f\"ur Theoretische Festk\"orperphysik, Universit\"at
Karlsruhe, D-76128 Karlsruhe, Germany}

\maketitle\abstracts{We analyze electron transport
through relatively short coherent conductors in the
presence of Coulomb interaction. We evaluate the current-voltage 
characteristics of
such conductors taking into account the effect of an external environment.
Within our model, at  large conductances and low $T$ the conductance
is suppressed by a universal factor  which depends only on the type of the
conductor. We also argue that at $T=0$
the system ``scatterer+shunt'' can be either an insulator or a metal depending on
whether its total resistance is larger or smaller than  $R_Q=h/e^2\approx 25.8$
k$\Omega$. In a metallic phase the Coulomb gap is fully suppressed by quantum
fluctuations. 
}

\section{Introduction}

Discrete nature of the electron charge plays a crucial role 
in various phenomena in mesoscopic physics, causing, for instance,
Coulomb blockade of electron
tunneling in metallic junctions \cite{AL,SZ,GD} and shot noise in mesoscopic
conductors \cite{BB}. It is of particular importance to understand how
an interplay of charge discreteness, Coulomb interaction and coherent
scattering may affect electron transport in disordered conductors at
sufficiently low temperatures. 

Recently we argued \cite{GZ00} that 
the interaction term in the current-voltage characteristics of a (relatively
short) coherent conductor is controlled by the parameter
\begin{equation}
\beta=\frac{\sum_nT_n(1-T_n)}{\sum_nT_n},
\label{param}
\end{equation}
which is already well known in the theory of shot noise \cite{BB}. Here $T_n$
are the transmissions of conducting modes. A similar conclusion was also
reached in Ref. 6 in the limit of a single conducting mode, in which
case the parameter (\ref{param}) reduces to $\beta =1-T_1$. 

In this paper we extend our previous results \cite{GZ00} by considering
the effect of an external environment on quantum transport through a 
coherent scatterer in the presence
of Coulomb interaction.

\section{Quasiclassical Langevin equation}

As in Ref. 5 we will consider electron transport through an
arbitrary coherent scatterer between two big reservoirs. The scatterer
conductance without interactions is defined by the standard Landauer formula $
1/R=(2e^2/h)\sum_nT_n$.
Phase and energy relaxation are only allowed in the
reservoirs and not during scattering, i.e. 
the scatterer is shorter than both dephasing and inelastic
relaxation lengths $L_{\varphi}$ and $L_{in}$. Coulomb effects in
the scatterer region are described by an effective
capacitance $C$. We also assume that the
scatterer is attached to an external voltage source $V_x$ via a linear
impedance $Z_S(\omega )$.

In the limit of sufficiently large energies and/or at large scatterer
conductances $g=R_Q/R \gg 1$ it is convenient to describe the system by means 
of the quasiclassical Langevin equation approach. This equation
can be derived from the real time path integral technique
\cite{Albert,AES,GZ92}. For a coherent conductor to be considered here
this derivation was performed in Ref. 5. The effect of a linear external
impedance can be incorporated in the same way as it was done for
the case of tunnel junctions \cite{GZ92}. Combining the results
\cite{GZ00,GZ92} one arrives at the following Langevin equation
\begin{equation} \frac{C}{e}\ddot\varphi + \frac{1}{eR}\dot\varphi
+\int \hat Z_S^{-1}(t-t')\frac{\dot \varphi (t')}{e}dt'-\frac{V_x}{Z_S(0)}=
\xi_1\cos\varphi +\xi_2\sin\varphi +\xi_3+\xi_S.
\label{langevin}
\end{equation}
Here $\dot \varphi (t)/e=V(t)$ is the fluctuating voltage across the conductor.
Eq. (\ref{langevin}) is sufficient provided $eV$ and
$T$ are  smaller than the typical inverse traversal time $1/\tau_{trav}$
(e.g. the Thouless energy in the case of diffusive conductors). Below we
will mainly address the limit $RC > \tau_{trav}$, however some of our results 
should remain valid in the opposite case as well. The
terms in the right-hand side of (\ref{langevin}) account for the current
noise. They  are defined by the correlators  \begin{equation}
\langle|\xi_{1}|_\omega^2\rangle =\langle|\xi_{2}|_\omega^2\rangle
=\frac{\beta}{ R}\omega\coth\frac{\omega}{2T},\;\;\;\;\;\;
\langle|\xi_{3}|_\omega^2\rangle = \frac{1-\beta}{
R}\omega\coth\frac{\omega}{2T}, \label{noise1} \end{equation} \begin{equation}
\langle|\xi_{s}|_\omega^2\rangle =
{\rm Re}\left( \frac{\omega }{Z_S(\omega )}\right)\coth\frac{\omega}{2T}.
\label{noise2}
\end{equation}

If we set $\varphi (t)=eV_xt$ and define the total fluctuating current 
$\delta I(t)=\xi_1\cos eV_xt+\xi_2\sin eV_xt +\xi_3,$
we will immediately reproduce the standart result \cite{Kh,BB}: 
\begin{equation}
\langle|\delta I|^2_\omega\rangle
=\frac{1}{R_Q}\biggl\{2\omega\coth\frac{\omega}{2T}\,\sum_n T_n^2
+\left[\sum_{\pm}(\omega \pm eV_x)\coth\frac{\omega \pm eV_x}{2T}\right] \sum_n T_n(1-T_n)\biggr\}.
\nonumber
\end{equation}
The $I-V$ curve for a conductor can be obtained by averaging eq.
(\ref{langevin}) over noise. We find
\begin{equation} IR=V-\langle [\xi_1\cos\varphi +\xi_2\sin\varphi ]\rangle .
\label{IV}
\end{equation}
The last term in eq. (\ref{langevin}) describes the effect of Coulomb
interaction. We note that this term depends only on the two stochastic
variables $\xi_1$ and $\xi_2$. Since the correlation functions for both
these variables (\ref{noise1}) are proportional to the parameter
$\beta$, the magnitude of the whole interaction term in (\ref{IV}) should
scale with the same parameter. Thus, the result (\ref{IV}) takes the form 
\begin{equation} R\frac{dI}{dV}=1-\beta f(V,T),  
\label{univ}
\end{equation}
where $f(V,T)$ is the universal function which depends on $R$ and $Z_S(\omega
)$. This function was already evaluated in the case of tunnel junctions
\cite{GZ92}. Defining $1/Z(\omega )=1/R-i\omega C +1/Z_S(\omega )$ and
proceeding perturbatively in Re$Z$, at $T \to 0$ one finds (cf., e.g., Refs. 9 and 6)   
\begin{equation}  \frac{d^2I}{dV^2}=\frac {e^2
\beta}{\pi RV}{\rm Re}[Z(eV)].
\label{Zarb}
\end{equation}
In a special case of a linear Ohmic environment
$Z_S(\omega ) \simeq R_S$ we find 
\begin{eqnarray}  f(V,T)&=&\frac{e}{\pi
}\int\limits_0^{+\infty}dt\,\frac{(\pi T)^2}{\sinh^2(\pi Tt)} {\rm
e}^{-F(t)}(1-{\rm e}^{-\frac{t}{R_0C}})\sin[eVt], 
\label{vax} \\
F(t)&=&-\frac{1}{g_0^2}\int_{-\infty}^{+\infty}dt'\frac{(\pi
T)^2}{\sinh^2(\pi Tt')}
\left(\beta g\cos[eVt']+(1-\beta )g+g_S\right)\hspace{0.7cm}
\nonumber\\
&&\times \left[|t'-t|-|t'|+R_0C\left({\rm e}^{-|t'-t|/R_0C}-{\rm
e}^{-|t'|/R_0C}\right) \right].
\label{F}
\end{eqnarray}
where $g_0\equiv R_Q/R_0=g+g_S$ and $g_S=R_Q/R_S$. Eqs. (\ref{vax}), (\ref{F})
work well provided either $g_0 \gg 1$ or max$(T,eV) \gg E_C$. In the limit
$g_0 \gg 1$ and max$(eV,T)  \gg g_0E_C\exp (-g_0/2)$ one can set $\exp (-F(t))
\simeq 1$. Then eqs.  (\ref{univ}), (\ref{vax}) yield the result 
\begin{equation} I=\frac{V}{R}-e\beta T{\rm Im}\left[
w\Psi\left(1+\frac{w}{2} \right)-iv\Psi\left(1+\frac{iv}{2} \right)
\right].
\label{iv}
\end{equation}
where $\Psi (x)$ is the digamma function, $w=u+iv$, $u=g_0E_C/\pi^2T$ 
and $v=eV/\pi T$.
At $T \to 0$ from (\ref{iv}) we obtain 
\begin{equation}
R\frac{dI}{dV}=1-\frac{\beta}{g_0}\ln\left(1+\frac{1}{(eVR_0C)^2}
\right),
\label{dif}
\end{equation}
while in the limit $eV/E_C \gg$ max$(1,g)$ we find $RI=V-\beta e/2C$. 
At $V \to 0$ from (\ref{iv}) we get 
\begin{equation}
f(0,T)=\frac{2}{g_0}\left[\gamma
+1+\ln\left(\frac{g_0E_C}{2\pi^2 T} \right) \right], \;\;\;\gamma \simeq
0.577. \label{intT}
\end{equation} 
The above logarithmic dependencies on $eV$ and $T$ should also hold for $R_0C
<\tau_{trav}$, in which case in eqs. (\ref{dif})
and (\ref{intT}) with the logarithmic accuracy one can set $g_0E_C \to 1/\tau_{trav}$.

At very low temperatures and voltages max$(T,eV) < g_0E_C \exp (-g_0/2)$
the interaction correction to the conductance of a coherent scatterer {\it
saturates} at a universal value which does not depend on the interaction but
only on the transmission distribution. This result follows
immediately from Eqs. (\ref{vax}), (\ref{F}).
Evaluating (\ref{F}) at long times we find $F(t) \simeq (2/g_0)(\ln
(t/R_0C)+\gamma )$ and performing the integral in (\ref{vax}), in the
leading order in $\beta /g_0  \ll 1$ we obtain  
\begin{equation} 
G=\frac{1-\beta}{R}=\frac{2e^2}{h}\sum_nT_n^2,
\label{lowT}
\end{equation}
This formula successfully reproduces a complete Coulomb blockade of
tunneling  $G \to 0$ in the limit $\beta \to 1$ (tunnel junctions),
demonstrates the absence of it for ballistic scatterers
($\beta \to 0$), and yields suppression of the Landauer
conductance by the factor 2/3 for diffusive conductors. 

\section{Comparison with experiments}
In order to compare our theoretical results with
recent experiments we will use the data reported recently by two groups
\cite{Weber,Krup}.

\begin{figure}[!ht]
\begin{center}
\psfig{figure=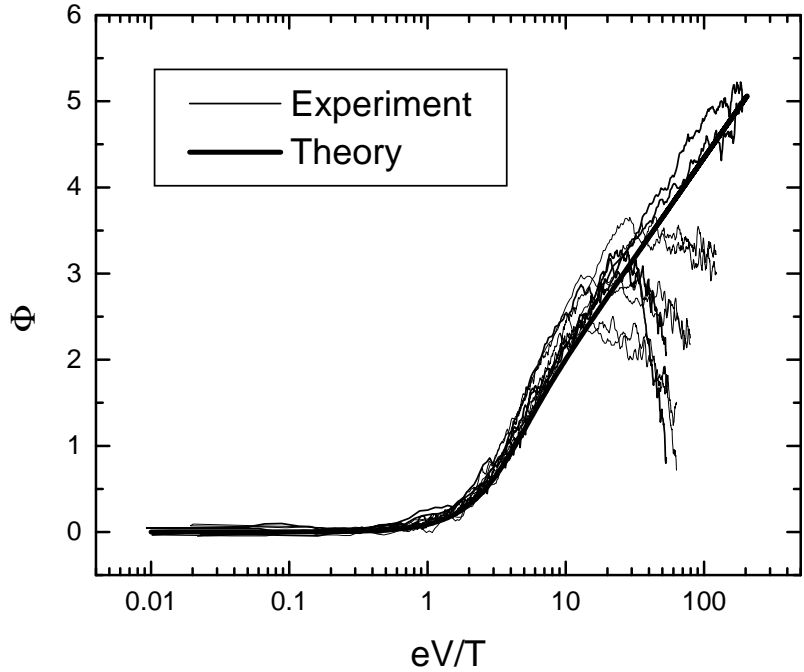,height=3.8in}
\caption{The function (\protect\ref{heiko}) plotted together with the experimental
data \protect\cite{Weber} for the sample No 1. Different experimental curves
correspond to different temperatures.
\label{fig:heiko}} 
\end{center}
\end{figure}

Weber {\it et al.} \cite{Weber} experimentally 
investigated the $I-V$ curve of a short ($L \sim 90$ nm $\ll
L_{\varphi},L_{in}$) diffusive conductor fabricated as a bridge between two
big metallic reservoirs. Even though the conductance of this bridge was large
$g \approx 2000$ the Coulomb blockade effect on the $I-V$ curve was
clearly visible becoming more pronounced with decreasing temperature, see Fig.
3 of Ref. 11. It was observed that the experimental data \cite{Weber} were
well described by the formula
\begin{equation}
G(V,T)=G(0,T_0)+A\ln (T/T_0) + A \Phi (eV/T),
\label{heiko1}
\end{equation}
where $A$ was determined to be $A\approx (0.4\div 0.7)\times R_Q^{-1}$
depending on the sample and $\Phi (eV/T)$ was  found to be a universal function
of $eV/T$ which tends to zero at $V \to 0$. The logarithmic dependence of
$G(0,T)$ on temperature (\ref{heiko1}) agrees with our eqs. (\ref{univ}),
(\ref{intT}). Both the function $\Phi (eV/T)$ and the value $A$ can easily be
evaluated within our theory. Making use of eq. (\ref{iv}) in the
experimentally relevant interval of temperatures and voltages we obtain 
$A=2\beta /R_Q$ and \begin{equation} \Phi (eV/T)={\rm Re}\left[
\Psi\left(1+i\frac{eV}{2\pi T}\right)+\gamma+i\frac{eV}{2\pi T}
\Psi'\left(1+ i\frac{eV}{2\pi T} \right)
\right].
\label{heiko}
\end{equation}
The function (\ref{heiko}) is plotted in Fig. 1 together with the
experimental data \cite{Weber} for the sample No. 1. One observes a very good
agreement between theory and experiment except at high voltages where heating
effects become dominant. From the same comparison for the sample No. 1 one finds $\beta \approx
0.25$ slightly smaller than the value $\beta =1/3$ 
expected in a diffusive limit. 

\begin{figure}[!ht]
\begin{center}
\psfig{figure=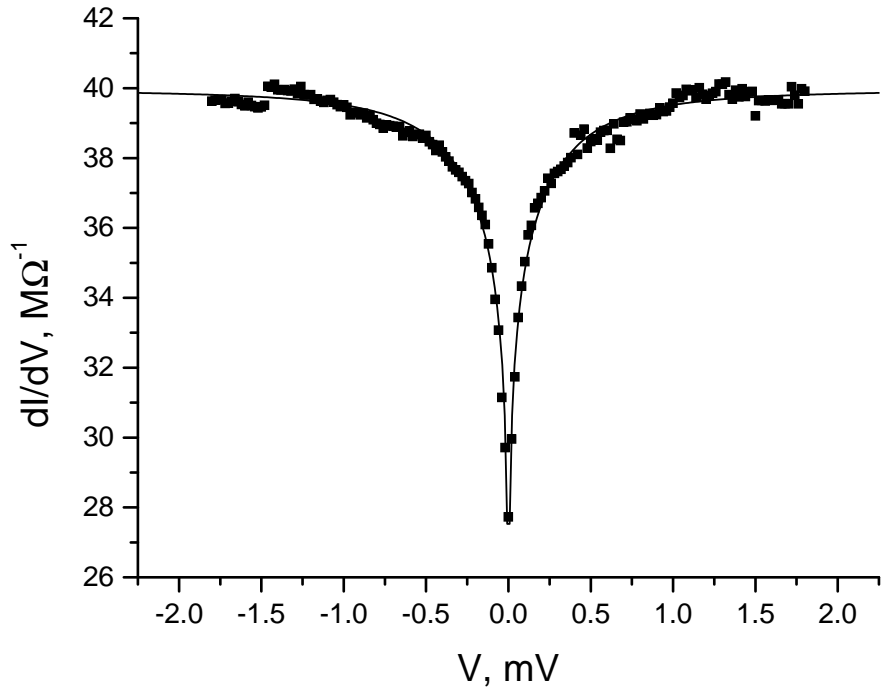,height=3.8in}
\caption{Eq. (\protect\ref{dif}) plotted together with the experimental data
\protect\cite{Krup} for the sample 1 with $R \protect\approx 2.5 $ K$\protect\Omega$ per square.
\protect\label{fig:sasha}} 
\end{center}
\end{figure}

In another experiment Krupenin {\it et al.} \cite{Krup} studied the $I-V$
curves of $Cr$ resistors fabricated in the form of 2d strips and also clearly
observed Coulomb blockade effects. The $\sim 1 \mu$m long samples \cite{Krup}
were not diffusive (their resistances were too large to be described by the
Drude formula) and most likely had a granular structure. For our comparison
we will adopt the model of a granular array with grains connected via
short coherent scatterers with some average value of the parameter $\beta$.
For the sample \cite{Krup} with the resistance 2.5 k$\Omega$ per square one can
estimate the average conductance of one scatterer as  $g \approx 10$. 
Again, a very good agreement between theory and experiment is observed, 
see Fig. 2. From the best fit one can extract the value $\beta \approx 0.35$.

A relatively low value of the scatterer conductance $g \approx 10$
reached in the experiments \cite{Krup} also allows to enter the regime
max$(T,eV) < gE_C \exp (-g/2)$, where the conductance was predicted to saturate at
the level  (\ref{lowT}). According to this prediction the ratio
between the values of $G$ measured in the limits of low and high voltages
should approach $1-\beta$. 
From the data \cite{Krup} we extract $\beta \approx 0.31$. 
This value is close to one
found from the fit in Fig. 2. Thus the data \cite{Krup} clearly support
both our theoretical predictions (\ref{dif}) and (\ref{lowT}). It is also
interesting that both estimates give the values of $\beta$ very close to
$\beta =1/3$ which one would expect for diffusive scatterers.

Finally, we notice that the results of recent conductance measurements
\cite{LTL} on multi-wall carbon nanotubes, with 
typical lengths  of order $L_{\varphi}$ or shorter, are also consistent
with the logarithmic dependence predicted in eq. (\ref{dif}). Further
experimental work on the subject would be highly desirable. 

\section{Instantons and metal-insulator phase transition}

Under which conditions does the result (\ref{lowT}) remain valid? One of
them was already formulated above: $g_0 \gg 1$, i.e. either $g$ or $g_S$ 
should be much larger than one. In this case the conductance $G$ saturates at
the value (\ref{lowT})  for max$(eV,T) < g_0E_C\exp (-g_0/2)$. In
order to establish another important limitation for the result (\ref{lowT})
let us recall that, even though our quasiclassical Langevin equation approach
does account for nonlinear in $\varphi$ effects, this approach may nevertheless
become insufficient at very low energies even for $g_0 \gg 1$ because
it does not include {\it instantons} \cite{pa91}, i.e. nontrivial saddle
points for the exact effective action.

An important part of the instanton analysis was carried out by Nazarov
\cite{Naz} who derived the renormalized Coulomb energy $\tilde E_C$
for a general coherent conductor within the exponential accuracy. The
result \cite{Naz} can be written in the form $\tilde E_C  \propto E_C\exp
(-ag)$. One can also go beyond the exponential accuracy and estimate the
pre-exponent in the expression for $\tilde E_C$. This was done in Ref. 16.
Without going into details here, we only quote the result \cite{GZ01}
\begin{equation}  
\tilde E_C/E_C \sim \left[\prod_{n}R_n \right]
\ln \left[\prod_{n}R_n^{-1}\right]   \sim ag\exp (-ag).  
\label{rengap}
\end{equation} 
This formula is valid for $ag \gg 1$, i.e. either at large
conductances $g \gg 1$ or, if $g \sim 1$, for very small values $R_n$
implying $a \gg 1$. 

The quantity $\tilde E_C$ (\ref{rengap}) plays
the role of an effective Coulomb gap in our problem. At $T <
\tilde E_C$ our Langevin equation analysis is insufficient and,
hence, eq. (\ref{lowT}) becomes inaccurate. In this regime the conductance
decreases with $T$ and the system is an insulator at $T=0$.

Let us now take into account the effect of an external linear Ohmic environment
$Z_S \simeq R_S$. In this case quantum fluctuations of the charge in the
shunt resistor will affect the Coulomb gap $\tilde E_C$ and -- as we shall see
-- may even lead to its total suppression provided $R_S$ is sufficiently low.

In order to proceed we will 
follow the analysis \cite{pa91,GZ01}. Treating the charge $q$ as a quantum
variable \cite{SZ} and integrating out the phase $\varphi $ one can map
our problem onto that of a linearly damped quantum particle $q$ in a periodic
potential.  This is a well-known problem \cite{s,SZ} which can be
treated, e.g., by means of the renormalization group (RG) technique.
Successively reducing the high frequency cutoff $\omega_c$ and integrating out
charges with higher frequencies one arrives at the standard RG equations
\cite{s}. After trivial manipulations these equations can be rewritten   
directly in terms of the combination $ag = \sum_n \ln R_n^{-1} $. One obtains
\cite{GZ01}
\begin{equation} 
d (ag) /d (\ln \omega_c)=(1-g_{\Sigma})(1+1/ag), \;\;\;\;\;\;\;d g_{\Sigma } /d
(\ln \omega_c) =0,
\label{ren2} 
\end{equation}
where $g_{\Sigma}=gg_S/g_0$ is the total dimensionless conductance. 
Eqs. (\ref{ren2}) are valid as long as $ag \gg 1$. One observes that
for $g_{\Sigma} <1$ the quantity $ag$ {\it decreases} in the course of 
renormalization. Hence, in that case the Coulomb gap remains nonzero,
the charge $q$ is localized and the system is an insulator at $T=0$. The
effective Coulomb gap $\tilde E_C$ can be defined as the energy scale at which
the renormalized value $ag$ becomes of order one. Then from (\ref{ren2}) one
finds \begin{equation}
\tilde E_C \sim E_C [ag\exp
(-ag)]^{\frac{1}{1-g_{\Sigma}}}\;.
\end{equation}
On the other hand, for $g_{\Sigma} > 1$ the combination $ag$ always
scales to {\it larger} values. In this case the Coulomb gap is 
fully suppressed by quantum fluctuations, $\tilde E_C=0$, the charge $q$
is delocalized and the
conductance remains nonzero (\ref{lowT}) even at $T=0$. This is a metallic
phase.  A quantum phase transition between the insulating and metallic phases
occurs at $g_{\Sigma}=1$.

Finally, let us briefly discuss possible implications of our
results for recent experiments \cite{Kr} which strongly indicate the presence
of a metal-insulator phase transition in various 2d disordered systems.
One can consider a (sufficiently small) coherent scatterer
with the dimensionless conductance $g$ viewing all other scatterers in the
system as an effective environment with the conductance $g_S$. {\it Assuming}
this environment to be Ohmic at sufficiently low frequencies, one immediately
arrives at the conclusion about the presence of a quantum  metal-insulator
phase transition  at $g_{\Sigma}$=1. In 2d systems one has $g \sim g_S \sim
g_0 \sim g_{\Sigma }$. Therefore in such systems this phase transition should
be expected at conductances $\sim 1/R_Q$, exactly as it was observed in many
experiments \cite{Kr}. Local properties of the insulating and metallic
phases are expected to be very different. In the insulating phase charges
should be localized around inhomogeneities (puddles) due to Coulomb blockade,
while in the metallic phase the Coulomb gap is suppressed and the charge
distribution should be much more uniform. These expectations are fully
consistent with experimental observations \cite{Ya}. Thus, there might be a
direct relation between the experimental results \cite{Kr,Ya} and the old
problem of a dissipative quantum phase transition \cite{s}.

\end{document}